\documentclass[sigconf]{acmart}
\AtBeginDocument{%
  }

\copyrightyear{2025}
\acmYear{2025}
\setcopyright{cc}
\setcctype{by-nc-sa}
\acmConference[ISWC '25]{Proceedings of the 2025 ACM International Symposium on Wearable Computers}{October 12--16, 2025}{Espoo, Finland}
\acmBooktitle{Proceedings of the 2025 ACM International Symposium on Wearable Computers (ISWC '25), October 12--16, 2025, Espoo, Finland}\acmDOI{10.1145/3715071.3750418}
\acmISBN{979-8-4007-1481-8/2025/10}

\begin{document}

\title{Beyond Symbols: Motion Perception Cues Enhance Dual-Task Performance with Wearable Directional Guidance}


\author{Qing Zhang}
\orcid{0000-0002-0622-932X}
\affiliation{%
  \institution{The University of Tokyo}
  \city{Tokyo}
  \country{Japan}}
\email{qzkiyoshi@gmail.com}

\author{Junyu Chen}
\orcid{0000-0002-9344-350X}
\affiliation{%
  \institution{The University of Tokyo}
  \city{Tokyo}
  \country{Japan}}
\email{924342513@qq.com}

\author{Yifei Huang}
\orcid{0000-0001-8067-6227}
\affiliation{%
  \institution{The University of Tokyo}
  \city{Tokyo}
  \country{Japan}}
\email{hyf015@gmail.com}

\author{Jing Huang}
\orcid{0000-0003-3128-2912}
\affiliation{%
  \institution{Tokyo University of the Arts}
  \city{Tokyo}
  \country{Japan}}
\email{hkoukenj@gmail.com}

\author{Thad Starner}
\orcid{0000-0001-8442-7842}
\affiliation{%
  \institution{Georgia Institute of Technology}
  \city{Atlanta}
  \country{USA}}
\email{thad@gatech.edu}

\author{Kai Kunze}
\orcid{0000-0003-2294-3774}
\affiliation{%
  \institution{Keio University}
  \city{Yokohama}
  \country{Japan}}
\email{kai@kmd.keio.ac.jp}

\author{Jun Rekimoto}
\orcid{0000-0002-3629-2514}
\affiliation{%
  \institution{The University of Tokyo}
  \city{Tokyo}
  \country{Japan}}
\affiliation{%
  \institution{Sony CSL Kyoto}
  \city{Kyoto}
  \country{Japan}}
\email{rekimoto@acm.org}

\renewcommand{\shortauthors}{Qing Zhang et al.}
\begin{abstract}
    Directional cues are crucial for environmental interaction. Conventional methods rely on symbolic visual or auditory reminders that require semantic interpretation, a process that proves challenging in demanding dual-tasking scenarios. We introduce a novel alternative for conveying directional cues on wearable displays: directly triggering motion perception using monocularly presented peripheral stimuli. This approach is designed for low visual interference, with the goal of reducing the need for gaze-switching and the complex cognitive processing associated with symbols. User studies demonstrate our method's potential to robustly convey directional cues. Compared to a conventional arrow-based technique in a demanding dual-task scenario, our motion-based approach resulted in significantly more accurate interpretation of these directional cues ($p=.008$) and showed a trend towards reduced errors on the concurrent primary task ($p=.066$).

\end{abstract}

\begin{teaserfigure}
  \includegraphics[width=\textwidth]{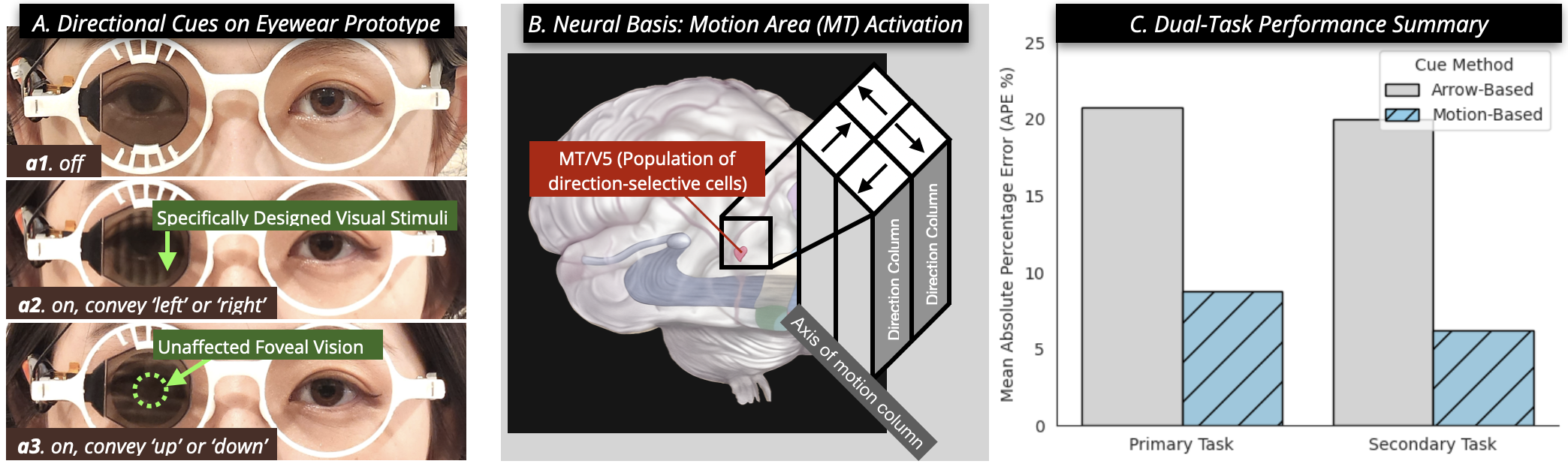}
    \caption{Overview of the proposed motion-based directional cueing system and its evaluation. (A) The wearable prototype in operation, showing how moving bar stimuli are displayed while leaving foveal vision unaffected. (B) Conceptual illustration of the neural basis, where peripheral stimuli are designed to directly activate the brain's motion-selective area (MT/V5) \cite{albright1984direction, mcgillBRAINFROM}. (C) Performance summary from the dual-task study, showing our motion-based cues resulted in lower mean Absolute Percentage Error (APE) compared to conventional arrow-based cues.}
  \Description{Enjoying the baseball game from the third-base
  seats. Ichiro Suzuki preparing to bat.}
  \label{fig:teaser}
\end{teaserfigure}

\maketitle

\section{Introduction}

    Directional cues are vital for daily interactions. Current methods often rely on visual symbols (e.g., arrows \cite{googleglass2013,narzt2006augmented, reitmayr2004collaborative, rekimoto1997navicam,amft2015making,bulling2016eyewear}) or language-based auditory signals that involve explicit semantic interpretation prior to action. Although effective in many situations, this cognitive processing step, which Licklider highlighted as time spent `getting into a position to think' \cite{licklider1960man}, can make such approaches error-prone and distracting, particularly in dual-tasking scenarios. This paper explores an alternative designed to mitigate these challenges. In accordance with principles of Calm Technology \cite{weiser1996designing} and effective wearable computing \cite{starner2001challenges}, interfaces should offer information in a subtle way, integrate seamlessly, and minimize cognitive load \cite{costanza2006eye}, which motivates of exploring of new cueing paradigms.

    Previous explorations of peripheral cues on head-worn displays \cite{gutwin2017peripheral, narzt2006augmented,nakao2016smart} have often faced challenges with visual intrusiveness, field-of-view obstruction, and maintaining efficacy under cognitive stress; for example, luminous elements can be distracting \cite{olwal20181d, konrad2016novel}, and interpreting peripheral symbols can add cognitive load \cite{newsome1988selective}. More recent advancements in Optical See-Through Head-Mounted Display (OHMD) interfaces focus on utilizing paracentral \cite{janaka2022paracentral, cai2023paraglassmenu} and near-peripheral vision \cite{chaturvedi2019peripheral, costanza2006eye} to present secondary information—such as progress updates \cite{janaka2022paracentral} or interactive menus \cite{cai2023paraglassmenu, poppinga2012ambiglasses}—with greater subtlety and reduced distraction from primary tasks \cite{ishiguro2011peripheral, mcatamney2006examination}. While these contemporary approaches effectively reduce gaze shifts and acknowledge the limitations of far-peripheral vision for detailed symbolic interpretation \cite{rosenholtz2016capabilities, strasburger2011peripheral}, they still fundamentally rely on the wearer perceiving and interpreting visual forms (e.g., text, icons, graphical shapes \cite{ku2019peritext}). This symbolic processing can be challenging under high cognitive load or when rapid reactions are paramount. The conveyance of basic directional cues through such methods often retains this layer of symbolic understanding, a specific challenge our work aims to address.

    This study introduces a novel method for conveying directional cues: directly triggering motion perception via monocularly presented stimuli. This approach aims to provide cues with low visual interference by design, potentially allowing the non-stimulated eye an unobstructed view of the environment. We hypothesize that this method will enable more direct, robust directional cue delivery, bypassing the necessity of gaze-switching, cognitive capture for symbolic understanding, and sophisticated optical design, particularly in dual-task situations. Our research contributes by: (1) introducing a novel approach for conveying directional cues that directly triggers motion perception via a monocular display; (2) characterizing the perceptual experience (e.g., accuracy, subjective comfort) of these monocular motion stimuli under various physical parameters (14 participants); and (3) investigating the potential for supporting visual dual-tasking by leveraging distinct information processing channels (10 participants).
    
    \section{Related Works}
    \label{sec:related_works}
        
        Our approach is grounded in principles of human visual processing and aims to address persistent challenges in designing intuitive cues for head-worn displays.

    \subsection{Leveraging Visual Perception for Directional Cues}    
        The human visual system has specialized mechanisms for motion perception, with directionally selective cells in areas such as the visual cortex (V1) and the middle temporal area (MT/V5) crucial to interpreting movement \cite{albright1984direction, maunsell1983functional, navarro2009optical, taylor2003new, vaney2012direction}. Motion perception can be used to convey directional information. Furthermore, foveal vision excels at detailed tasks, while peripheral vision is highly attuned to motion and can process information with less focused attention \cite{bartram2003moticons, rosenholtz2016capabilities, strasburger2011peripheral}. Peripheral pathways are suitable for unobtrusive secondary cues. Our system utilizes monocular presentation, allowing the contralateral eye an unobstructed view of the environment. This setup was used to characterize the stimulus's perceptual qualities (User Study 1), and the role of binocular mechanisms like summation or rivalry in the overall perceptual experience remains an area for deeper investigation \cite{blake1981further, park2008anthropometry}.

    \subsection{Directional Cues and Information Presentation on OHMDs}
        Conveying information effectively on Optical See-Through Head-Mounted Displays (OHMDs) without overwhelming the user or disrupting primary tasks is a significant HCI challenge. Much research has focused on presenting secondary information, such as notifications or system status, often exploring paracentral and near-peripheral visual fields to minimize distraction from a central task (e.g., social interaction or IoT control) \cite{janaka2022paracentral, cai2023paraglassmenu}. These systems employ various visual forms, such as progress bars or interactive menus with icons and text.

        Prior work on computational eyewear has also explored peripheral LEDs or abstract visual patterns for notifications and cues \cite{costanza2006eye, olwal20181d, poppinga2012ambiglasses, sutton2022look, shenchromagazer}. While these aim for subtlety, they often still rely on the user learning and interpreting a symbolic mapping (e.g., a specific light pattern means `turn left') or processing static visual forms. Specific attempts at directional cues on HMDs have included arrow-based symbols or other visual indicators \cite{feiner1997touring, livingston2002augmented}. However, these approaches can suffer from issues like gaze shifting \cite{walter2022low}, cognitive overhead for symbol interpretation, or visual intrusiveness, especially under cognitive load \cite{suzuki2024measuring}. The challenge of dual-tasking with visual displays is well-documented, often leading to performance degradation due to attention capture and the serial nature of focused visual processing \cite{appel2019smartphone, foyle1993attentional, gish1995human}.
        
        Our work differs by focusing specifically on conveying directional cues (up, down, left, right) by directly engaging motion perception pathways, with the intention of reducing the cognitive steps required for the interpretation of symbols.

\section{System Design: A Wearable Modality for Motion-Based Cues}
\label{sec:system_design}

Building on perceptual principles (Section~\ref{sec:related_works}), we designed and implemented a lightweight, head-worn eyewear system \cite{starner2001challenges} to deliver directional cues. The system aims to provide gaze-independent, non-symbolic information by directly triggering motion perception, thereby aiming to minimize the reliance on semantic interpretation or dedicated foveal attention. Our approach leverages the distinct characteristics of foveal and peripheral vision \cite{janaka2022paracentral, jung2018ensuring} and uses a monocular display to present stimuli whose perceptual qualities (e.g., noticeability, comfort in various physical contrast settings) were characterized (User Study 1). Conceptually, our method involves three key aspects: (1) \textbf{Activating Motion Perception Directly:} Triggering directional motion using specifically designed moving patterns presented peripherally to stimulate directionally selective pathways (Section~\ref{sec:related_works}). (2) \textbf{Minimizing Visual Interference:} Perceptually managing visual interference by presenting the stimulus to one eye and  allowing the contralateral eye an unobstructed view of the environment (elaborated in Section~\ref{sec:stimulus_design}),  characterized in User Study 1. (3) \textbf{Separating Visual Tasks:} Enabling foveal vision to remain focused on a primary task while peripheral vision processes directional cues.

    \begin{figure}
        \centering
        \includegraphics[width=0.8\linewidth]{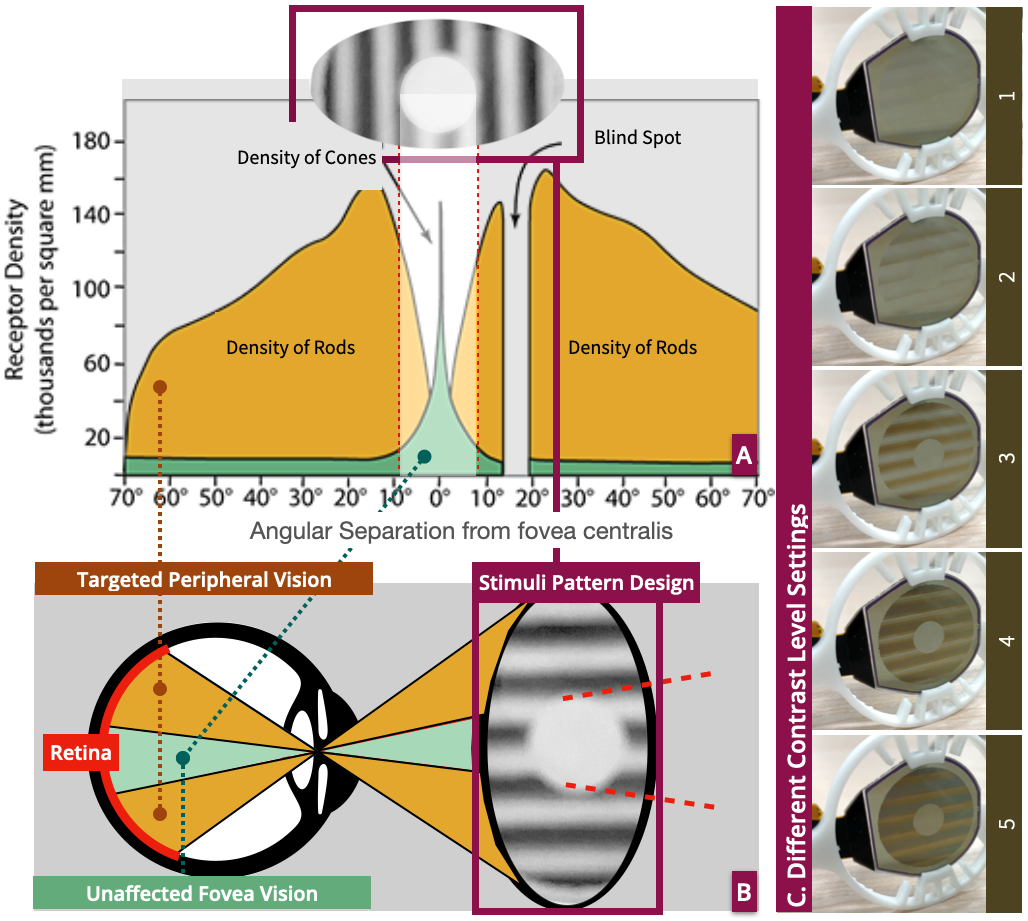}
            \caption{Stimulus design targeting peripheral photoreceptor distribution. (A) Density of cones (foveal, high acuity) and rods (peripheral, motion-sensitive) across the retina, adapted from Osterberg \cite{retina1935distribution}. (B) Our stimulus pattern targets the rod-dense periphery for motion cues while the foveal region remains unobscured for primary visual tasks. (C) Examples of the stimulus pattern appearance at the five different physical contrast levels evaluated in User Study 1.}
        \label{fig:distribution}
    \end{figure}

\subsection{Stimulus Design for Triggering Motion Perception}
\label{sec:stimulus_design}

\begin{figure}
        \centering
        \includegraphics[width=0.8\columnwidth]{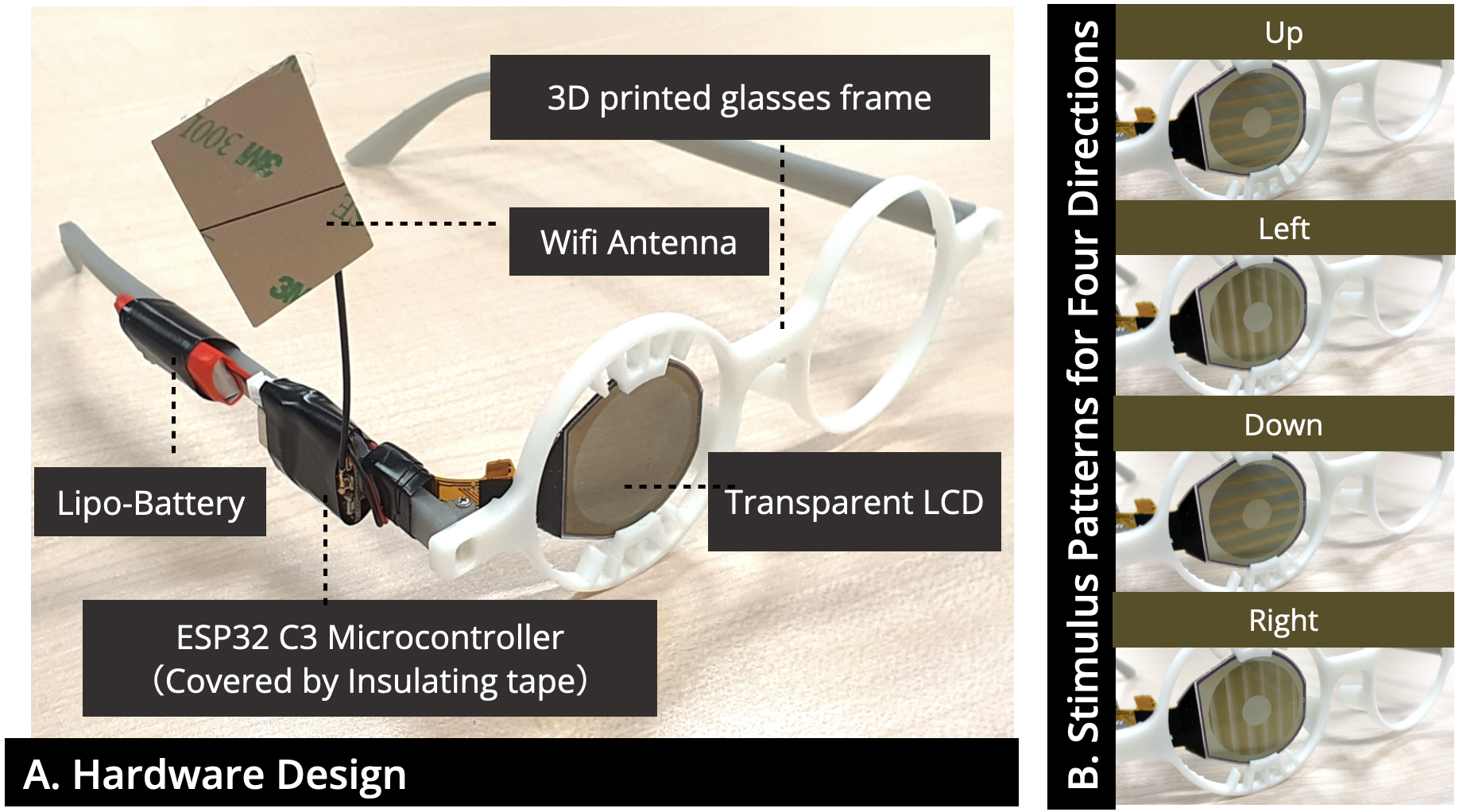}
           \caption{Wearable prototype and stimulus visualization. (A) Hardware design, detailing the 3D-printed frame, transparent Liquid Crystal Display (LCD) integrated into the right lens holder, ESP32-C3 microcontroller, LiPo battery, and Wi-Fi antenna. (B) Representative frames illustrating the moving bar patterns displayed on the transparent LCD to convey the four cardinal directional cues.}

        \label{fig:prototype_directions}
\end{figure}

Instead of conventional symbols, our modality relies on specifically designed moving patterns presented peripherally to evoke a perception of motion in the intended direction (examples in Figure~\ref{fig:prototype_directions}B).

\textbf{Pattern Choice and Rationale.} We use high-contrast, low spatial frequency horizontal or vertical bars moving across the display (Figure~\ref{fig:distribution} and \ref{fig:prototype_directions}B). This choice is intended to effectively stimulate directionally selective neurons (e.g., in retinal Directionally Selective Ganglion Cells (DSGCs) and cortical area MT/V5) \cite{maunsell1983functional, barlow1965mechanism}, a choice intended to engage motion-sensitive pathways and thus potentially require less semantic interpretation than traditional symbols. This leverages peripheral vision's superior motion detection capabilities compared to foveal vision \cite{strasburger2011peripheral} and has the potential for parallel processing without significantly disrupting foveal tasks \cite{janaka2022paracentral, Leibowitz1984}.

\textbf{Peripheral Targeting \& Foveal Sparing.} The stimuli are designed to primarily engage the peripheral visual field while preserving foveal vision. This approach is informed by the distinct functions and photoreceptor distributions across the retina (Figure~\ref{fig:distribution}), targeting the high density of motion-sensitive rods in the periphery \cite{provis2013adaptation}. The fovea, crucial for high-acuity tasks, remains unobstructed by a software-defined transparent central aperture, corresponding to a 20-pixel radius in display settings (found through experimentation). This adheres to the principles of assigning tasks to appropriate visual channels to support dual-tasking \cite{foyle1993attentional, katsuyama1989effects}.

\textbf{Stimulus Parameters.} We refined key parameters iteratively through pilot tests involving five individuals to balance motion perception intensity and visual comfort:
\textit{Contrast:} High contrast (black bars on a transparent background) is used to maximize the motion signal.
\textit{Speed:} A speed of 3 pixels per frame was found to reliably trigger motion perception without being reported as too distracting or too slow to be missed.
\textit{Spatial Frequency \& Bar Spacing:} Bars are 20 pixels wide/high, with a spacing of 8 pixels between bars (see Figure~\ref{fig:prototype_directions}B for the pattern design). This spacing was determined after multiple iterations to ensure sufficient signal intensity. 

\textbf{Monocular Presentation: }The stimulus is presented to only one eye. This design choice simplifies the prototype and ensures the contralateral eye maintains an unobstructed view of the real-world scene. This setup enables the characterization of the monocular stimulus's perceptual qualities across different physical contrast levels, as demonstrated in User Study 1. Future work could investigate how specific binocular phenomena—such as summation and rivalry—contribute to these perceptual effects \cite{blake1981further, alais2005binocular}.

\subsection{Wearable Prototype Implementation}
\label{sec:prototype}

To demonstrate and evaluate this interaction modality, we developed a wireless, lightweight monocle prototype (Figure~\ref{fig:prototype_directions}A).

\textbf{Hardware Components.} A single circular transparent monochrome LCD (gy128128, 1.6-inch diameter, 128x128 resolution, ~20g total weight including custom printed circuit board) is mounted in the right lens holder of a custom 3D-printed resin glasses frame (interpupillary distance approx. 63mm). An ESP32-C3 microcontroller drives the display and handles wireless communication, powered by a 3.7V 140mAh LiPo battery for mobility.

\textbf{Ergonomics and Optics.} The transparent LCD allows users to see through the display. It is positioned close to the eyeball (approximately 10 mm, adjustable via nose pads/temples). This proximity forces/encourages users to look \textbf{through} the display at their environment rather than focusing \textbf{on} the display surface itself, aligning with peripheral stimulation goals and leveraging the human visual system's difficulty on focusing on very near objects \cite{duane1922studies, navarro2009optical, zhang2022seeing, zhang2022gazesync}. This design refers to suggestions that placing cues in locations consistently within the user's peripheral field of view leads to faster reactions \cite{harrison2009locate}. 
Although ocular dominance can influence visual tasks \cite{coren1977fifty, reiss1997ocular}, we did not examine its effects in this study, which focuses on single-eye presentation.  Dominance advantages are not significant consistently  across all contexts \cite{handa2005effects, mccarley2004visual}; however, it is a good area for future work here.

\begin{figure}
        \centering
        \includegraphics[width=0.75\columnwidth]{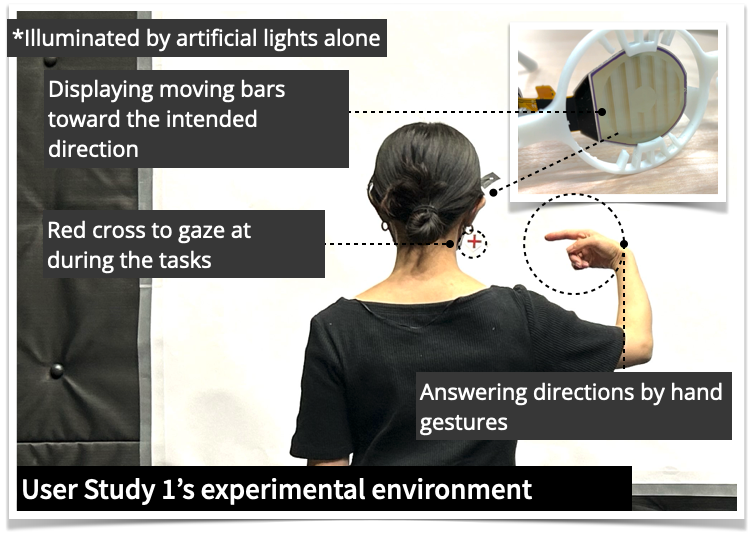}
            \caption{User Study 1 experimental environment. Participants faced a height-adjusted central fixation cross on a white background (approx. 40cm distance) while wearing the prototype.}
        \label{fig:study1_setup}
\end{figure}

\section{User Study 1: Accuracy of Directional Perception in a Controlled Environment}

Study 1 assesses the baseline accuracy of perceiving directions (Up, Down, Left, Right) using the proposed modality under controlled laboratory conditions with varying stimulus contrast levels.

Fourteen participants (9 female, 5 male, aged 23 to 33, mean 25.9, SD 3.4) participated in the study. They stood facing a white background with a central fixation cross (similar to Figure~\ref{fig:study1_setup}). Calibration involved adjusting the prototype’s nose pads and temples for a secure fit, then aligning the display’s central transparent aperture with the participant's line of sight to the fixation cross. Wearing the calibrated prototype, participants experienced moving bar patterns corresponding to the four cardinal directions, presented twice each in random order for each of the five contrast levels. Participants indicated the perceived direction (e.g., via hand gestures or verbal report), and the experimenter recorded the accuracy. To assess the impact of the hardware contrast setting on participants' subjective experience during this accuracy task, we also analyzed perceived contrast and difficulty ratings collected after each block. Participants were required to rate the perceived contrast and the perceived difficulty level in distinguishing the directional cues.

\subsection{User Study 1: Result}

\begin{figure}
    \centering
    \includegraphics[width=0.85\linewidth]{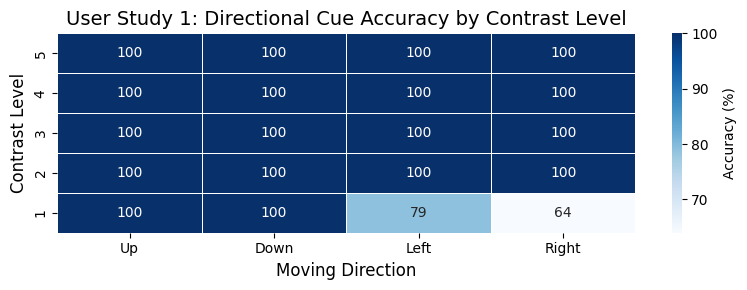}
    \caption{Accuracy of directional cue perception in User Study 1 (N=14) across five contrast levels and four motion directions. Cells are annotated with the mean accuracy percentage for each condition. Accuracy was 100\% in most cases, with a notable decrease for horizontal motion at the lowest contrast level.}
    \label{fig:user_study_1_heatmap}
\end{figure}

\textbf{Accuracy.} The modality demonstrated high accuracy in this controlled setting. Accuracy reached 100\% for levels 2-5 (Figure~\ref{fig:user_study_1_heatmap}). At contrast level 1, accuracy remained high for vertical motion (Up/Down: 100\%) but was slightly lower for horizontal motion (Left: 79\%, Right: 64\%).

\begin{figure}
    \centering
    \includegraphics[width=\linewidth]{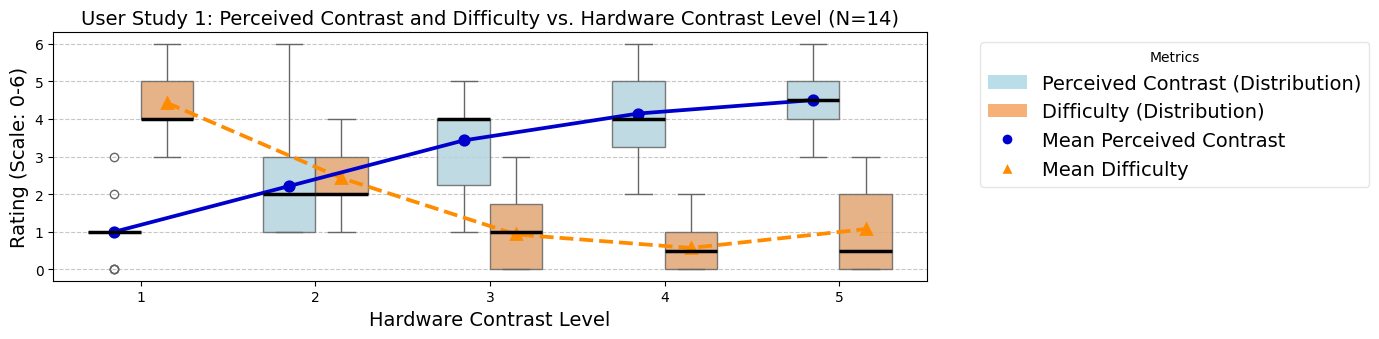}
    \caption{User Study 1: Subjective ratings vs. physical contrast (N=14). Box plots show perceived contrast (blue) and difficulty (orange) across five contrast levels, with medians shown as thick black lines. Lines connecting the means illustrate that as hardware contrast increases, perceived contrast increases and perceived difficulty decreases.}
    \label{fig:user_study_1_plot}
\end{figure}

\textbf{Subjective Ratings.} A Kruskal-Wallis H test was chosen as it is a non-parametric method suitable for ordinal rating data that may not meet the normality assumptions required for a parametric ANOVA. The test revealed a statistically significant effect of hardware contrast level on perceived contrast ratings ($H(4)=40.63,p<.0001$). Post-hoc comparisons using the Dunn-Bonferroni test showed that perceived contrast for Level 1 was significantly lower than for Levels 3, 4, and 5 (all $p < .01$). Similarly, a statistically significant effect was found for perceived difficulty ratings ($H(4)=44.33,p<.0001$). Post-hoc analysis confirmed that perceived difficulty for Level 1 was significantly higher than for Levels 3, 4, and 5 (all $p < .001$), consistent with the accuracy results (Figure~\ref{fig:user_study_1_plot}).

\subsection{User Study 1: Discussion}

The high accuracy achieved in this study establishes the fundamental feasibility of conveying directional cues using our proposed modality in a controlled setting. Performance remained remarkably robust even at low contrast levels. User Study 1 characterized the relationship between physical stimulus contrast and participants' subjective ratings of perceived contrast and difficulty under the monocular presentation setup. The post-hoc analysis of these ratings pinpointed a clear perceptual threshold, with the lowest contrast setting being rated as significantly more difficult and less clear than the higher settings. This result aligns with the known sensitivity of peripheral vision to motion \cite{gutwin2017peripheral} and the potential for direct activation of directionally selective pathways \cite{taylor2003new, snowden_2012_basic174}, allowing effective perception even with subtle or low-salience stimuli. Participant feedback frequently mentioned ``feeling'' the direction without necessarily seeing the pattern clearly, further supporting the possibility of perception occurring without focused attention or detailed visual recognition.

The slightly lower accuracy for horizontal motion compared to vertical motion at the lowest contrast level is noted and may warrant further investigation regarding potential anisotropies in peripheral motion perception for this specific stimulus type. This performance drop corresponds directly with the subjective ratings, where participants found this lowest contrast level significantly more difficult to interpret. Overall, however, the results demonstrate the potential of this approach to provide accurate directional information under baseline conditions.

        \begin{figure}
            \centering
            \includegraphics[width=0.85\columnwidth]{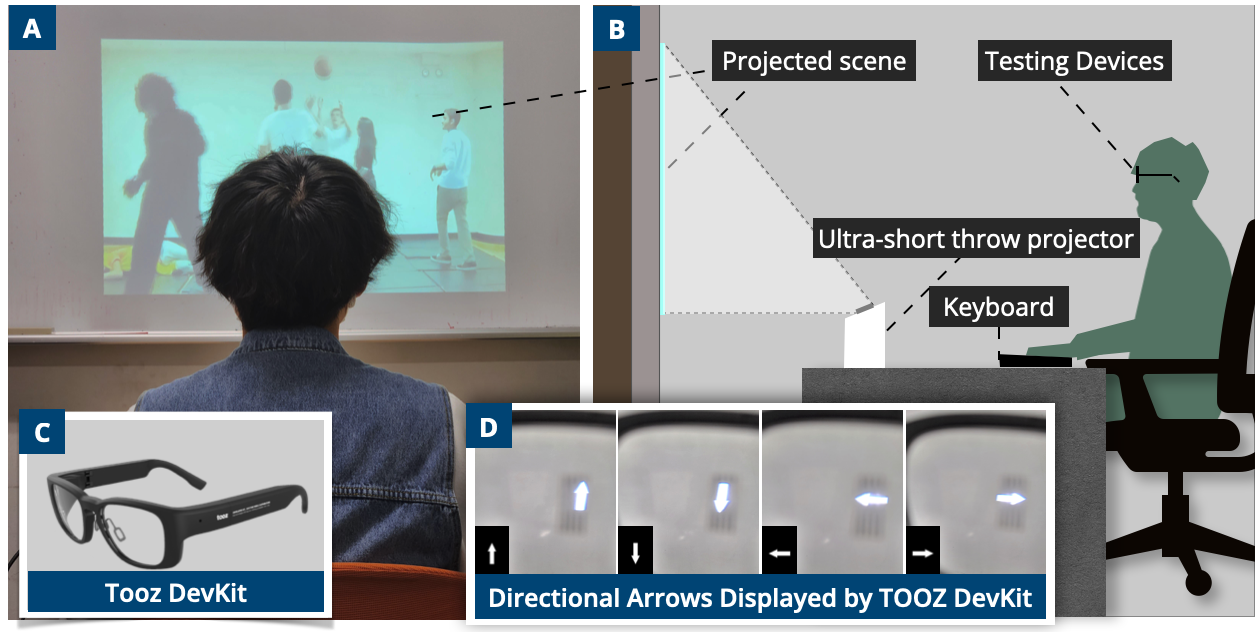}
                \caption{User Study 2 dual-task experimental setup. (A) Participant view of the primary task: a projected video for the Selective Attention Test. (B) Diagram illustrating the setup: participant using one of the ``Testing Devices'' (either our prototype or the Tooz DevKit) and a keyboard for responses, while viewing the video projected by an ultra-short throw projector. (C) The Tooz DevKit used for displaying the arrow-based cues. (D) Visual examples of the directional arrow cues (up, down, left, right) as displayed by the Tooz DevKit during the arrow-based condition.}

            \label{fig:dual_task_env}
        \end{figure}

\begin{table*}[htbp]
        \centering
        \footnotesize
    \begin{tabular}{|c|c|c|c|c|c|c|c|c|c|c|c|}
        \hline
        \multicolumn{2}{|c|}{} & P1 & P2 & P3 & P4 & P5 & P6 & P7 & P8 & P9 & P10 \\
        \hline
        Arrow-based method  & Pass Counting & $!$\textcolor{red}{35}/25 & 23/25 & 22/25 & 20/25 & 23/25 & $!$\textcolor{red}{40}/25 & 19/25 & 18/25 & 24/25 & $!$\textcolor{red}{26}/25 \\
        \cline{2-12}
        in OST AR glass& Directional Answers  & $\Delta7(2)/8$ & 6/8 & 7/8 & 3/8 & 5/8 & 6/8 & 6/8 & 5/8 & 3/8 & \textbf{8/8} \\
        \hline
        \textbf{Our Approach} & Pass Counting & 23/25 & \textbf{25/25} & 22/25 & \textbf{25/25} & 24/25 & 22/25 & 17/25 & 22/25 & 23/25 & \textbf{25/25} \\
        \cline{2-12}
        & Directional Answers  & \textbf{8/8} & \textbf{8/8} & \textbf{8/8} & 5/8 & \textbf{8/8} & \textbf{8/8} & $\ast6(1)/8$ & \textbf{8/8} & 7/8 & \textbf{$\ast8(1)/8$} \\
        \hline
        
        \end{tabular}
        \caption{User Study 2: Performance comparison between our approach and an arrow-based method using an OST-AR glass (Tooz). The `Arrow-based method' and `Our Approach' are arranged sequentially in the table for comparison purposes only and do not imply the order in which the methods were tested. The order was counterbalanced during the test to control for order effects. Note: ($!$) indicates that the answer exceeds the correct answer (25). ($\bigtriangleup$) denotes incorrect responses; e.g., ``$\Delta7(2)/8$'' means the participant pressed the arrow key 9 times, with 7 correct responses and 2 incorrect responses. ($\ast$) indicates a false positive, where a participant responded to a stimulus that was not actually present. These infrequent events were only observed to occur immediately following a real  ``left'' or ``right'' stimulus.}

            \label{tab:user_study_2_RawData}
\end{table*}

        \begin{figure}
            \centering
            \includegraphics[width=\columnwidth]{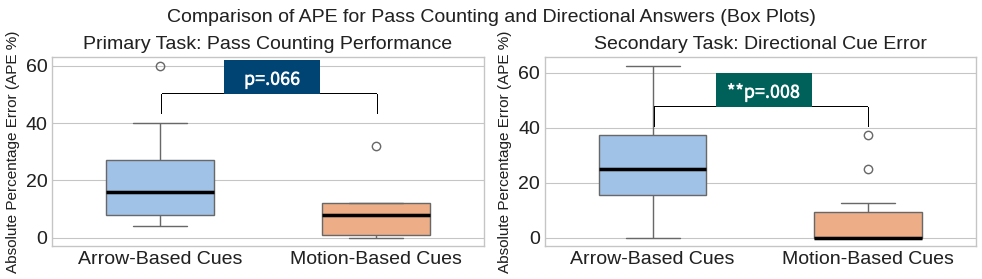}
                \caption{User Study 2: Comparison of Absolute Percentage Error (APE) for the primary task (Pass Counting, left) and secondary task (Directional Answers, right) between our Motion-Based Cues and conventional Arrow-Based Cues. Lower APE indicates better performance. Boxes show interquartile range (IQR), and whiskers extend to 1.5x IQR.}

            \label{fig:user_study_2_result}
        \end{figure}

\section{User Study 2: Evaluating Cues in a Dual-Task Scenario}
\label{sec:user_study_2}

To assess performance under cognitive load, this study compared our motion-based directional cues with a conventional arrow-based approach during a demanding dual-task. This experiment addresses concerns about visual notification efficacy in realistic contexts \cite{costanza2006eye} and explores the use of peripheral vision for secondary information \cite{Leibowitz1984, janaka2022paracentral}, contrasting direct motion perception with symbolic comprehension.

\textbf{Participants and Apparatus: }Ten participants (6 female, 4 male, aged 23-35, M=26.8, SD=3.9) from User Study 1 continued in this within-subject study. 
For the arrow-based condition, we used a Tooz DevKit to display white (RGB: 255,255,255) arrow symbols (up, down, left, right), designed in a style common to AR/VR/HUD systems (see Figure~\ref{fig:dual_task_env}D). For our motion-based approach, stimuli were presented at contrast level 4, based on User Study 1 findings balancing efficacy and comfort.

\textbf{Task and Procedure: }Participants performed a modified Selective Attention Test (a variant of the ``Invisible Gorilla'' experiment \cite{simons1999gorillas}) as the primary cognitive load task. They counted basketball passes made by a team in black in a 53-second video, projected onto a whiteboard (due to our prototype's polarized LCD). 

Concurrently, participants received directional cues (either from our prototype or the Tooz DevKit arrows) and responded by pressing corresponding arrow keys on a wired keyboard. The devices and the presentation order of the cueing methods were counterbalanced. Figure~\ref{fig:dual_task_env} illustrates the experimental environment.

\subsection{User Study 2: Results}
Table~\ref{tab:user_study_2_RawData} and Figure~\ref{fig:user_study_2_result} detail the performance comparison.

\textit{Primary Task (Pass Counting) Performance:} Our motion-based approach led to a lower mean Absolute Percentage Error (APE) in the primary pass-counting task. The mean APE for pass counting was 8.8\% with our method, compared to 20.8\% for the arrow-based approach (Figure \ref{fig:user_study_2_result}, left). The arrow-based method also resulted in more instances of significant overestimation (e.g., P1, P6, P10 in Table \ref{tab:user_study_2_RawData}). To account for the small sample size and potential non-normality of the data, we used the Wilcoxon signed-rank test. The difference in APE between the motion-based cues (Mdn = 8.0\%) and arrow-based cues (Mdn = 16.0\%) was not statistically significant ($W = 6.5, p = .066$).

\textit{Secondary Task (Directional Cue) Accuracy:} Our motion-based approach yielded significantly higher accuracy in responding to directional cues. The mean APE for directional answers was 6.25\% with our method, versus 20\% for the arrow-based approach (Figure \ref{fig:user_study_2_result}, right). A Wilcoxon signed-rank test confirmed that the APE for directional answers was significantly lower with our motion-based approach (Mdn = 0.0\%) compared to the arrow-based method (Mdn = 25.0\%) ($W = 0.0, p = .008$). With our method, most participants achieved perfect or near-perfect scores. Some misinterpretations and extra key presses were noted with the arrow-based method. A few instances of perceiving non-existent motion were recorded with our approach (Table \ref{tab:user_study_2_RawData}, asterisks).

\textit{Participant Insights:} Post-task interviews revealed that arrows often diverted attention from the primary task. One participant noted, ``\textit{whenever the arrows suddenly appeared, my attention was instantly diverted... I would add a few counts to account for what I believed I might have missed}.'' Another `` \textit{mistakenly counted the arrows as passes.}'' For our method, occasional opposite-direction responses for horizontal movements were reported.

\subsection{User Study 2: Discussion}

The results from User Study 2 indicate that our motion-perception-based cues were interpreted significantly more accurately than conventional arrow symbols in this dual-task scenario. Furthermore, while the reduction in error for the primary cognitive task (pass counting) did not achieve statistical significance ($p=.066$), the mean performance was notably better with our motion-based cues (mean APE: 8.8\% vs. 20.8\%; median APE: 8.0\% vs. 16.0\%). This trend suggests that our approach may interfere less with a cognitively demanding primary task.

The lower mean APE in pass counting with our approach suggests a trend towards reduced cognitive load compared to the arrow-based method. Participants reported the arrow-based method as more disruptive, leading to attentional switching and errors in the primary task. The arrow-based method's higher APE variability also indicates less consistent performance across users.

The statistically significant superior accuracy in identifying directions with our method ($p=.008$) strongly supports its potential for effective cueing during concurrent visual tasks. The poorer performance of the arrow-based approach is likely attributable to the need for foveal attention and gaze shifts to resolve arrowheads, especially when not directly fixated, causing them to appear visually similar and leading to missed primary task events. In contrast, our design, which preserves unaffected foveal vision and uses specifically designed monocular stimuli, likely contributed to its robust performance in the secondary task and the observed trend of improved primary task performance. The reported instances of perceiving non-existent motion with our approach, though infrequent, highlight the need for future work on optimizing stimulus parameters or implementing context-aware activation to minimize such perceptual artifacts.

\section{General Discussion and Conclusion} 
\label{sec:discussion_conclusion}

Our findings demonstrate that directly triggering motion perception via monocularly presented peripheral stimuli is a viable and effective method for conveying directional cues. User Study 1 established the high accuracy of this approach across various physical contrast levels and characterized the perceptual experience of these novel cues. Crucially, our dual-task study revealed that our method led to significantly more accurate cue interpretation while also showing a strong trend towards reducing errors on the primary task. This result suggests a promising potential for reducing overall cognitive burden compared to traditional symbolic cues.

The significantly improved accuracy in cue interpretation and the trend towards reduced primary task interference with our symbol-free, motion-based approach have implications for the design of future wearable interfaces, particularly where minimizing cognitive load associated with cue processing and reducing gaze shifts is paramount (e.g., navigation, assistive technologies, high-workload environments). It suggests a promising alternative to traditional symbolic cues on OHMDs. However, our exploration has limitations. The current prototype was tested in controlled settings with specific stimuli; real-world environmental factors and a wider range of user characteristics (e.g., varying sensitivities to motion) warrant further investigation. The occasional perception of non-existent motion in User Study 2, though infrequent, also points to the need for refining stimulus parameters or exploring context-aware activation.

Future work should explore a broader design space for motion stimuli, investigate adaptation effects over longer-term use, and test the approach in more ecologically valid scenarios. Different input modalities for acknowledging cues could also be explored. In conclusion, this research provides compelling evidence for a novel, perception-driven method for conveying directional information on wearable displays, offering a pathway towards more intuitive and less cognitively demanding human-computer interaction.

\begin{acks}
This work was supported by JST Moonshot R\&D Grant JPMJMS2012, JSPS Grant-in-Aid for Early-Career Scientists JP25K21241, and JPNP23025 commissioned by the New Energy and Industrial Technology Development Organization (NEDO).
\end{acks}

\bibliographystyle{ACM-Reference-Format}
\bibliography{sample-base}


\end{document}